# Multiple Cross-Layer Design Based Complete Architecture for Mobile Adhoc Networks

R.Venkatachalam
Research Scholar
K.S.R College of Technology, Tiruchengode-637 215
Namakkal District, Tamilnadu, India

Dr.A.Krishnan
Dean
K.S.R College of Engineering, Tiruchengode-637 215
Namakkal District, Tamilnadu, India

*Abstract*— **Different cross-layer design for mobile adhoc network focuses on different optimization purpose, different Quality of Service (QoS) metric and the functions like delay, priority handling, security, etc. Existing cross-layer designs provide individual solution for congestion control, fault tolerance, power conservation, energy minimization and flow control and the major drawback is of high cost and overhead. In this paper, we propose to design multiple cross-layer design based architecture to provide a combined solution for link failure management, power conservation, congestion control and admission control. By simulation results, we show that the average end-to-end delay, average energy consumption and the packet loss are considerably reduced with the increase in high throughput and good delivery ratio.**

*Keywords: Cross-Layer; MANETs; end-to-end; Packet loss; Delivery ratio; congestion control*

## I. INTRODUCTION

Ad-hoc networks are multi-hop wireless networks where all nodes cooperatively maintain network connectivity. These types of networks are useful in situation where temporary network connectivity is needed. A mobile ad hoc network (MANET) [1] is a group of mobile, wireless nodes which cooperatively and spontaneously form a network independent of any fixed infrastructure or centralized administration. Though the major motivation of studying ad hoc networks comes from military usage, they will also be useful in any form of tactical communications such as disaster recovery, explorations, law enforcements, and in various forms of home and personal area networks. In order to provide communication throughout the network, the mobile nodes must cooperate to handle network functions, such as packet routing.

Routing is the most active research field in mobile ad hoc networking. Minimizing the number of hops is no longer the objective of a routing algorithm, but rather the optimization of multiple parameters such as packet error rate over the route, energy consumption, network survivability, routing overhead, route setup and repair speed, possibility of establishing parallel routes, etc. Many routing protocols for mobile ad-hoc network have appeared recently [2] - [9].

A critical issue for MANETs is that the activity of nodes is power-constrained. Developing routing protocols for MANETs has been an extensive research area during the past few years. In particular, energy efficient routing may be the most important design criteria for MANETs since mobile nodes will be powered by batteries with limited capacity.

In order to achieve the desired vertical optimization goal, the useful information is inter-communicated by the different layers of the network protocol stack which is considered as cross-layer or inter-layer networking. The requirements of the quality of service may vary with applications and hence the network or higher layers function should directly rely on the information from the lower physical and MAC layers. Different layers can share locally available information by using interlayer interaction. This will significantly improve the performance.

There are many cross-layer designs for different optimization purpose. Different cross-layer design focuses on different optimization purpose, different QoS metric, one or more of the followings: delay, priority handling, security, etc. Obviously every system needs more than one cross-layer design to achieve overall QoS optimization.

### A. Problems of the Existing System Architecture of Cross-layer Designs

- Only the local link information from its MAC layer is used by the congestion avoidance algorithm. The local information is inadequate to replicate the network situation if the whole network is unstable.

- In general the cross-layer designs involve the combination of layers physical-MAC-network, MAC-network, Network-Transport only. But, there is no work made on complete integration of MAC-Network-Transport layers.

- The cross-layer designs provide individual solution for congestion control, fault tolerance, power conservation, energy minimization and flow control. There is no complete and combined solution for the above issues.

- Expensive and High Overhead

In this paper, we propose to design multiple cross-layer based designs architecture to provide a combined solution for link failure management, power conservation, congestion control and admission control.

The paper is organized as follows. Section 2 presents the related work done. Section 3 presents a detailed description of our proposed architecture. Section 4 presents the simulation results and conclusion is given in Section 5.

## II. RELATED WORK

Tom Goff et al. [11] have investigated adding proactive route selection and maintenance to on-demand ad-hoc routing algorithms. More specifically, when a path was likely to be broken, a warning has sent to the source indicating the likelihood of a disconnection. The source was then initiate path discovery early, potentially avoiding the disconnection altogether.

Hong-Peng Wang and Lin Cui [12] have discussed that the need for an efficient routing protocol in mobile ad hoc network has widely proclaimed. Their work has presented an enhanced AODV protocol, a scheme to make mobile nodes more aware of the local connectivity to its neighbors in the network. Their scheme has extended the original HELLO message in AODV but with lower overhead. At the same time it has prevented the potential unidirectional links in the network to some extent.

PremaLatha et al. [13] have discussed that in mobile ad hoc wireless networks, multiple mobile stations has communicated without the support of a centralized coordination station for the scheduling of transmissions. Their study has deal a combination of medium access control procedure employing distributed coordination function and suitable transport layer mechanism which has improved QoS guarantee in Transport layer. In their proposed method, IEEE802.11e and Adaptive Increase Multiplicative Decrease (AIMD) mechanism have been combined to analyze the quality of service in cross layer.

RamaChandran and ShanmugaVel [14] have discussed that in fourth generation (4G) wireless networks and beyond, it is strongly anticipated that mobile ad hoc networks are used to economically extend their coverage and capacity. In their work, they have proposed and studied three cross-layer designs among physical, medium access control and routing (network) layers, using Received Signal Strength (RSS) as cross-layer interaction parameter for energy conservation, unidirectional link rejection and reliable route formation in mobile ad hoc networks.

Lijun Chen et al. [15] have considered jointly optimal design of cross-layer congestion control, routing and scheduling for ad hoc wireless networks. They have first formulated the rate constraint and scheduling constraint using multi-commodity flow variables, and formulate resource allocation in networks with fixed wireless channels as a utility maximization problem with these constraints. By dual decomposition, the resource allocation problem has naturally decomposed into three sub problems: congestion control, routing and scheduling that interact through congestion price.

Xinsheng Xia et al. [16] have introduced a method for cross-layer design in mobile ad hoc networks. They have used fuzzy logic system (FLS) to coordinate physical layer, data link layer and application layer for cross-layer design. Ground speed, average delay and packets successful transmission ratio are selected as antecedents for the FLS. The output of FLS has provided adjusting factors for the AMC (Adaptive Modulation and Coding), transmission power, retransmission times and rate control decision.

A.n. al-khwildi, S. khan et al [17] proposes a novel routing technique called Adaptive Link-Weight (ALW) routing protocol. ALW adaptively selects an optimum route on the basis of available bandwidth, low delay and long route lifetime. The technique adapts a cross-layer framework where the ALW is integrated with application and physical layer. The proposed design allows applications to convey preferences to the ALW protocol to override the default path selection mechanism.

B. RamaChandran and S. Shanmugavel [18] proposed a simple cross layer design between physical (PHY) and Medium Access Control (MAC) layers for power conservation based on transmission power control. The Carrier Sense Multiple Access with Collision Avoidance (CSMA/CA) mechanism of IEEE 802.11 Wireless Local Area Network (WLAN) standard is integrated with the power control algorithm. In this method, the exchange of RTS/CTS control signals is used to piggyback the necessary information to enable the transmitting nodes to discover the required minimum amount of power that is needed to transmit their data packets.

## III. PROPOSED MCBA DESIGN

*A. Overview*

In this paper, we propose to design multiple cross-layer based designs architecture to provide a combined solution for link failure management, power conservation, congestion control and admission control.

1) **Link Failure Management**: Using the received signal strength from physical layer, link quality can be predicted and links with low signal strength will be discarded from the route selection.

2) **Power Conservation:** Using the MAC layer RTS/CTS packets exchange, the minimum required power can be estimated and accordingly the sender can adjust its transmitting power.

3) **Congestion Control**: From the MAC layer, contention and channel interference of the nodes can be estimated and notified to the application layer. Based on these estimations, at the application layer, the transmission rate can be adjusted, to avoid congestion.

4) **Admission Control**: From the physical layer, the bandwidth capacity of the nodes can be estimated. Using this information, an admission control mechanism at the MAC layer, admits or rejects the flows according to their requested bandwidth.

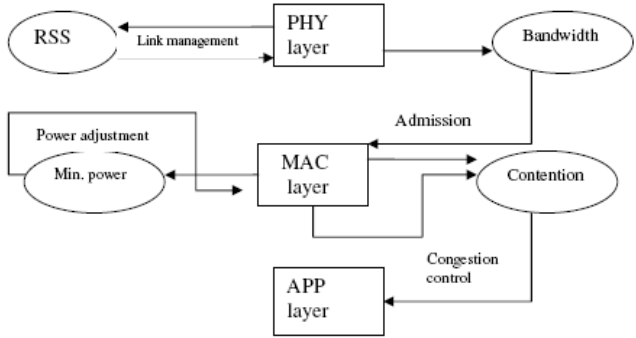

Figure.1 Cross-Layer Architecture

*B. Link Failure Management*

The link quality can be predicted by means of the received signal strength from physical layer and the links with low signal strength will be discarded from the route selection.

The received signal strength in cross layer designs is calculated at the physical layer and it can be accessed at the top layers as shown in the figure 1.The measured value of received signal strength will be transferred to the MAC layer along with the signal [15]. The procedures at physical layers have to be personalized. This value in MAC layer calculations is used if required or to pass the routing layers along with the routing control packets. This value is stored in the routing/neighbour tables and it is also used in some of the decision making process. As an interlayer interaction parameter, the received signal strength which is related to the physical layer is passed to the top layers. The received signal strength improves the performance of the mobile ad hoc networks by adjusting the medium access and routing protocols as per the required cross layer design,

The IEEE 802.11 is reliable MAC protocol. Since the received signal strength must reach every exposed node, it assumes the fixed maximum transmission power. When a sending node transmits RTS packet, it attaches its transmissions power. The receiving node measures the signal strength received for free–space propagation model while receiving the RTS packet [15].

$$P_R = P_T (\lambda/4\pi d)^2 G_T G_R \qquad (1)$$

Where $\lambda$ wavelength of carrier, $d$ is distance between sender and receiver. $G_T$ and $G_R$ are unity gain of transmitting and receiving omni directional antennas, respectively

*C. Power Conservation*

Using the MAC layer RTS/CTS packets exchange, the minimum required power can be estimated and accordingly the sender can adjust its transmitting power.

The receiving node calculates the path loss experienced as,

$$Path\ loss = P_T - P_R \qquad (2)$$

The node then calculates the minimum required transmission power ($P_{T\min}$) as,

$$P_{T\min} = k*(path\ loss + R_{TH}) \qquad (3)$$

Where $R_{TH}$ is receiver threshold, the minimum required power required for proper signal detection. The multiplication factor "$k$" is considered to provide marginal hike in minimum required transmission power to withstand against the effect of interference on packet reception.

In our cross-layer design, based on the type number at MAC layer the RREP packet is acknowledged and from PHY layer the received signal strength information is obtained and is passed to routing layer. Hence, the path loss experienced by the packet is calculated by the nodes that receives AODV's RREP packet and the minimum required transmission power is computed using the equations (2) and (3).The $P_{T\min}$ is stored in the routing table with the next hop against the destination. In order to get the minimum required transmission power in cross-layer design, the node sending RTS has to refer the routing table. It tunes it's transmit power to this value and also inserts this value in RTS as extra field so that the receiving node can tune to this power while sending its CTS packet. Subsequently by using the required minimum transmit power level, the data packet from the sender and ACK packet from the receiver can also be transmitted. This scheme clearly reflects the cross-layer interaction among PHY-MAC – Routing layers.

*D. Congestion Control*

From the MAC layer, contention and channel interference of the nodes can be estimated and notified to the application layer. Based on these estimations, at the application layer, the transmission rate can be adjusted, to avoid congestion.

In this network, we consider IEEE 802.11 MAC with the distributed coordination function (DCF). It has the packet sequence as request-to-send (RTS), clear-to-send (CTS), and data, acknowledge (ACK). The amount of time between the receipt of one packet and the transmission of the next is called a short inter frame space (SIFS). Then the channel occupation due to MAC contention will be

$$C_{occ} = t_{RTS} + t_{CTS} + 3t_{SIFS} \qquad (4)$$

Where $t_{RTS}$ and $t_{CTS}$ are the time consumed on RTS and CTS, respectively and $t_{SIFS}$ is the SIFS period.

Then the MAC overhead $OH_{MAC}$ can be represented as

$$OH_{MAC} = C_{occ} + t_{acc} \qquad (5)$$

Where $t_{acc}$ is the time taken due to access contention.

The amount of MAC overhead is mainly dependent upon the medium access contention, and the number of packet collisions. That is, $OH_{MAC}$ is strongly related to the congestion around a given node.

$OH_{MAC}$ can exceed threshold value $T_{rh}$, if congestion is incurred and not controlled, and it can dramatically decrease the capacity of a congested link.

The channel resource $\Delta S$ can be calculated as,

$$\Delta S = \frac{(T_{rh} - OH_{MAC})}{OH_{MAC}} \times S \qquad (6)$$

Where $S$ is the current traffic load

If $OH_{MAC} < T_{rh}$, then $\Delta S$ will be positive and

If $OH_{MAC} > T_{rh}$, then $\Delta S$ will be negative.

The transmission rate $rt$ is dynamically adjusted according to the explicit feedback $fd$. Namely,

$$rt = rt + fd \qquad (7)$$

If $\Delta S$ is positive, then the transmission rate $rt$ will be increased as,

$rt = rt + \Delta S$ and

if $\Delta S$ is negative, then the transmission rate $rt$ will be reduced as,

$rt = rt - \Delta S$

Thus the traffic rate is adaptively adjusted according to the MAC contention.

*E. Admission Control*

From the physical layer, the bandwidth capacity of the nodes can be estimated. Using this information, an admission control mechanism at the MAC layer, admits or rejects the flows according to their requested bandwidth.

In the bandwidth estimation method, the sender's current bandwidth usage as well as the sender's one-hop neighbors' current bandwidth usage is credited onto the standard "Hello" message. Each host estimates its feasible bandwidth based on the information provided in the "Hello" messages and knowledge of the frequency reuse pattern. This approach avoids creating extra control messages by using the "Hello" messages to disseminate the bandwidth information. Every host estimates its occupied bandwidth by scrutinizing the packets it provides into the network. A band width utilization register records the value at the host and updates periodically. We modify the "Hello" message to include two fields. The initial field includes host address, consumed bandwidth, timestamp, and the second field includes neighbors' addresses, consumed bandwidth, timestamp. The host receives a "Hello" message from its neighbors, and concludes whether this "Hello" is a restructured one by examining the message's timestamp.

Once a host knows the bandwidth consumption of its first neighbors and its second neighbors, the feasible bandwidth FBW is estimated as

$$FBW = (CHBW - UBW / WT) \qquad (8)$$

Where, $CHBW$ - channel bandwidth, $UBW$ - used or consumed bandwidth, $WT$ - weight factor, $RBW$ -required bandwidth, $MinBW$ -Minimum Bandwidth, $MaxBW$ - Maximum Bandwidth

**Algorithm**

*Step 1: If $FBW < RBW$, the source node will be rejected, else it further checks the required band width*

*Step 2: If $RBW < MinBW$, the flow can be admitted.*

*Step 3: If $RBW > MaxBW$, the flow is rejected.*

*Step 4: If $MinBW > RBW < MaxBW$, a probing packet is sent to the destination node to obtain the FBW at the destination.*

IV. SIMULATION MODEL AND PARAMETERS

We use NS2 to simulate our proposed protocol in our simulation, the channel capacity of mobile hosts is set to the same value: 2 Mbps. We use the distributed coordination function (DCF) of IEEE 802.11 for wireless LANs as the MAC layer protocol. It has the functionality to notify the network layer about link breakage.

In our simulation, 50 mobile nodes move in a 1500 meter x 500 meter rectangular region for 100 seconds simulation time. We assume each node moves independently with the same average speed. All nodes have the same transmission range of 250 meters. In our simulation, the speed is set as 5m/s. The simulated traffic is Constant Bit Rate (CBR). The pause time of the mobile node is varied as 0,10,20,30 and 40.

Our simulation settings and parameters are summarized in table I.

TABLE I. SIMULATION SETTINGS

| No. of Nodes | 25,50,75 and 100 |
|---|---|
| Area Size | 1500 X 500 |
| Mac | 802.11 |
| Radio Range | 250m |
| Simulation Time | 100 sec |
| Traffic Source | CBR |
| Packet Size | 512 |
| Mobility Model | Random Way Point |
| Speed | 5m/s |
| Pause time | 10 |
| Receiving Power | 0.395 W |
| Transmit Power | 0.660 W |
| Idle Power | 0.035 W |
| Initial Energy | 4.7 J |

*A. Performance Metrics*

We compare our MCBA protocol with the AOMDV [2] and AODV [6] protocols**.** We evaluate mainly the performance according to the following metrics, by varying the number of nodes as 25, 50, 75 and 100.

**Control overhead:** The control overhead is defined as the total number of routing control packets normalized by the total number of received data packets. It occurs while estimating the metrics for Link failure management, Power Conservation, Congestion control, Admission control and exchanging these metrics among different layers.

**Average end-to-end delay:** The end-to-end-delay is averaged over all surviving data packets from the sources to the destinations.

**Average Packet Delivery Ratio:** It is the ratio of the number of packets received successfully and the total number of packets sent

**Throughput:** It is the number of packets received successfully.

**Drop:** It is the number of packets dropped

**Average Energy:** It is the average energy consumption of all nodes in sending, receiving and forward operations

*B. Results*

Based On Number of Nodes

In this experiment, we vary the number of nodes as 25, 50, 75 and 100

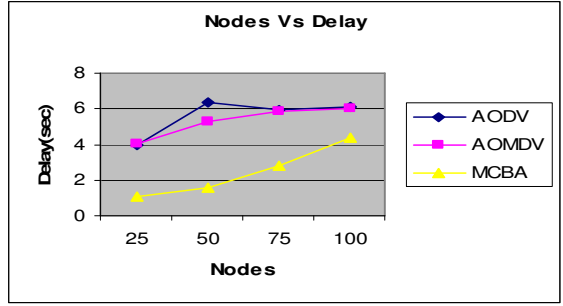

Figure 2. Nodes Vs Delay

Figure 2, represents that the average end-to-end delay of the proposed MCBA protocol is very less when compared to AOMDV and AODV protocol.

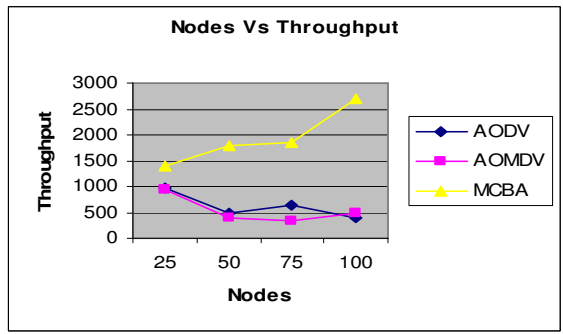

Figure 3. Nodes Vs Throughput

Figure 3 gives the throughput of all the protocols when the number of nodes is increased. As we can see from the figure, the throughput is more in the case of MCBA, than AOMDV and AODV.

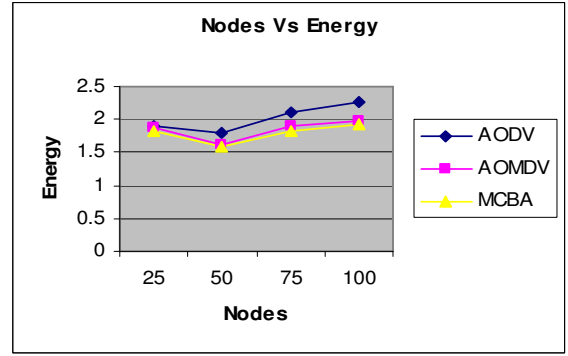

Figure 4. Nodes Vs Energy

Figure 4 shows the results of energy consumption. From the results, we can see that MCBA protocol has less energy than the AOMDV and AODV protocols, since it has the energy efficient routing

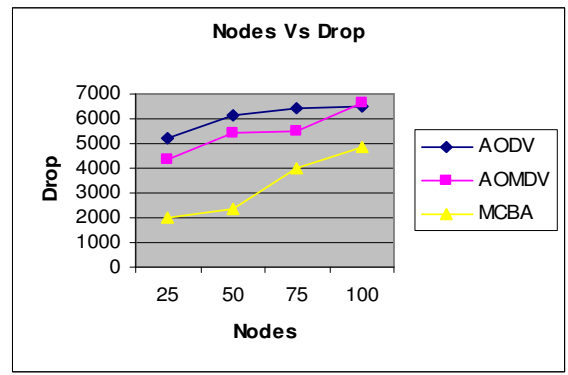

Figure 5. Nodes Vs Drop

Figure 5 ensures that the packets dropped in MCBA are less when compared to AODV and AOMDV.

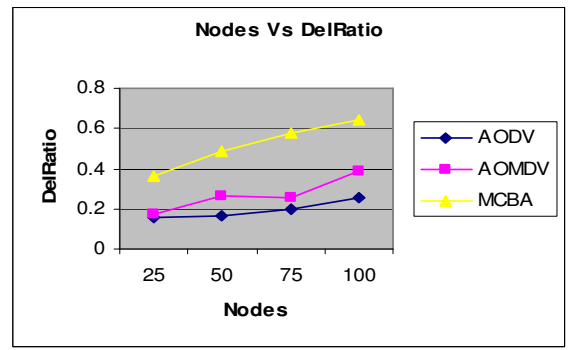

Figure 6. Nodes Vs DelRatio

Figure 6 presents the packet delivery ratio of both AODV and AOMDV protocols in comparison with MCBA protocol. Since the packet drop is less and the throughput is more,

MCBA achieves good delivery ratio, compared to AOMDV and AODV.

## V. CONCLUSION

In this paper, we have designed multiple cross-layer design based architecture to provide a combined solution for link failure management, power conservation, congestion control and admission control. The link quality can be predicted by means of the received signal strength from physical layer and the links with low signal strength will be discarded from the route selection. Using the MAC layer RTS/CTS packets exchange, the minimum required power can be estimated and accordingly the sender can adjust its transmitting power. From the MAC layer, contention and channel interference of the nodes can be estimated and notified to the application layer. Based on these estimations, at the application layer, the transmission rate can be adjusted, to avoid congestion. From the physical layer, the bandwidth capacity of the nodes can be estimated. Using this information, an admission control mechanism at the MAC layer, admits or rejects the flows according to their requested bandwidth. By simulation results, we have shown that the average end-to-end delay, average energy consumption and the packet loss are considerably reduced with the increase in high throughput and good delivery ratio.

AUTHORS PROFILE

**R.Venkatachalam** received the MCA degree from the Bharathidasan University, Tiruchirappalli, Tamilnadu, India, in 1995. He is currently a research scholar in K.S.R. College of Technology, Tiruchengode, Tamilnadu, India. He has passed UGC-NET exam in Computer Science and Applications in June 2006. He has published 7 papers in the national conference proceedings. His research interests include mobile adhoc networks and mobile communications. He is a life member of the Computer Society of India (CSI) and Indian Society for Technical Education (ISTE).

**Dr. A. Krishnan** received the PhD degree from the Indian Institute of Technology (IIT) Kanpur, India, in 1979. He is currently Dean, K.S.R. College of Engineering, Tiruchengode, Tamilnadu, India. His research interests include control systems and computer networks. He is a senior member of the IEEE, Computer Society of India (CSI) and Indian Society for Technical Education (ISTE).